\documentclass[prl,aps,twocolumn,superscriptaddress,showpacs,floatfix]{revtex4}
\usepackage{graphicx}
\usepackage{amsmath}

\begin{document}

\title{Effects of edge disorder in nano-scale antiferromagnetic clusters}

\author{Wanzhou Zhang }
\affiliation{Department of Physics, Beijing Normal University, Beijing 100875,  China}

\author{Wenan Guo} 
\email{waguo@bnu.edu.cn}
\affiliation{Department of Physics, Beijing Normal University, Beijing 100875, China}

\author{Ling Wang}
\affiliation{Department of Physics, Boston University, 590 Commonwealth Avenue, Boston, Massachusetts 02215, USA}
\affiliation{Institut der Theoretischen Physik, Universit\"at Wien, Boltzmanngasse 3, A-1090 Vienna, Austria}

\author{Kaj H. H\"oglund}
\affiliation{Department of Physics, {\AA}bo Akademi University, Porthansgatan 3, FI-20500 Turku, Finland}

\author{Anders W. Sandvik}
\email{sandvik@bu.edu}
\affiliation{Department of Physics, Boston University, 590 Commonwealth Avenue, Boston, Massachusetts 02215, USA}

\date{\today}


\begin{abstract}
We study the distribution of local magnetic susceptibilities in the two-dimensional antiferromagnetic $S=1/2$ Heisenberg model on 
various random clusters, in order to determine whether effects of edge disorder could be detected in NMR experiments (through the line shape, 
as given by the distribution of local Knight shifts). Although the effects depend strongly on the nature of the edge and 
the cluster size, our results indicate that line widths broader than the average shift should be expected even in clusters 
as large as $\approx 1000$ lattice spacing in diameter. Experimental investigations of the NMR line width should give insights 
into the magnetic structure of the edges.
\end{abstract}

\maketitle
Experimental, theoretical, and computational studies of two dimensional (2D) quantum spin systems have taken a prominent place 
in condensed matter physics during the past two decades. The $S=1/2$ Heisenberg model has successfully explained the bulk 
antiferromagnetism in the parent compounds of the high-$T_{\rm c}$ cuprates \cite{aeppli,chn}, and other quasi-2D materials 
\cite{ronnow, butcher}. It has also been realized that controlled studies of systems with impurities can give additional valuable 
information on the electronic structure and interactions in these strongly-correlated materials \cite{impexp}. Impurity effects are also 
of interest in their own right, as they reflect fascinating quantum phenomena not present in translationally 
invariant systems \cite{sachdev,hoglund1,metlitski1,liu}. 

In this Letter, we address the physics of defects beyond single-spin impurities in 2D antiferromagnets, by considering various 
forms of edge disorder. Prominent effects of effectively free chain ends are known in 1D systems, where detailed comparisons of
theories, computational model studies, and experiments are possible \cite{eggert,takigawa}. In contrast, 2D edges have not been paid 
much attention to, presumably because one would naively expect their influence to be relatively small in experiments probing spatially averaged 
properties, due to typically small edge-to-bulk ratios. With the increasing focus on physics on 
the nano-scale, it should also be interesting to investigate small antiferromagnetic clusters, where edge effects 
could dominate the physics. To this end, an initial quantum Monte Carlo (QMC) study of both smooth and rough edges in 
the 2D $S=1/2$ Heisenberg model was recently carried out \cite{hoglund2}. A suppression of the magnetic susceptibility 
at a smooth edge was found at low temperature ($T$), contrary to the naively expected enhancement due to the smaller 
coordination number of the edge spins. The edge contribution to the susceptibility 
is logarithmically divergent for $T\to 0$, as was later found also within a continuum field-theory description \cite{metlitski2}. 
Another interesting observation was a dimerization pattern at the edge, which can be seen as a precursor 
to valence-bond solid state that can be realized if additional interactions are included
\cite{jqmodel}. These effects may not be easy to observe experimentally, however, because edge roughness masks 
this behavior \cite{hoglund2}. The roughness introduces subsets of spins which are effectively weakly 
coupled to the bulk system, leading to a strongly enhanced susceptibility. The behavior reflects a complex, and still not well 
understood, interplay between geometric roughness at the microscopic scale and the collective, macroscopic behavior 
of interacting spins.

The nature of the clusters in powder samples of quasi-2D antiferromagnets such as La$_2$CuO$_4$ is currently not known precisely, and, 
to our knowledge, there have not been any efforts to investigate possible effects of finite-size clusters in, e.g., nuclear-magnetic-resonance 
(NMR) experiments. Cluster diameters $\agt 1000$ lattice spacings can be expected \cite{takigawacomm}. One would expect free edges 
to lead to broadening of the NMR line (a distribution of Knight shifts), as in 1D \cite{takigawa}. The question is, whether this broadening can 
be observed, and what information it can provide on the structure of the edges. 

We here present a systematic study of the Knight-shift distribution in finite $S=1/2$ Heisenberg 2D clusters constructed with varying 
amounts of edge roughness. We calculate the width of the distribution of local magnetic susceptibilities as a function of the cluster size. 
The results indicate that edge effects should be very prominent even for relatively large clusters, of width 100-1000 lattice spacings, at 
temperatures where the cuprates are paramagnetic (i.e., above the N\'eel temperature, $T_{\rm N}$, where order sets in due to weak 3D couplings 
or anisotropies).

With only on-site hyperfine couplings taken into account, the NMR Knight shift of a Cu nuclear spin at lattice site ${\bf r}$ is proportional 
to the local susceptibility $\chi_l({\mathbf r})$, which is given by
\begin{equation}
\chi_l({\mathbf r})=\beta \sum_i \langle S_i^z S_{\mathbf r}^z\rangle,
\label{chil}
\end{equation}
where $\beta=1/T$ (in units where $k_B=1$). In an infinite uniform system $\chi_l({\mathbf r})=\chi$; the bulk 
susceptibility. Disorder or open edges in a finite system lead to local variations and, thus, a broadened NMR line. 
In reality, in cuprates there are significant transferred hyperfine couplings also to nearest-neighbor Cu sites and the Knight shift should be 
modified accordingly \cite{millis}. Here, in this initial proof-of-concept study, we do not consider these couplings and just report results of 
the completely local susceptibility. The methods we use can, however, easily be extended to any NMR form factor. 

Distributions of susceptibilities due to isolated vacancies (corresponding to Cu substituted by Zn in cuprates) in the 2D 
Heisenberg model \cite{anfuso,alexander} and ladder systems \cite{alexander} have previously been studied using QMC simulations. 
Here we use similar techniques to study open-boundary clusters with various types of edges. We use the stochastic series 
expansion method with efficient loop updates and improved estimators \cite{sse,evertz} to evaluate (\ref{chil}) for various types of clusters 
at different temperatures.

We build three different types of clusters: A) As in Ref.~\cite{hoglund2}, starting from an open 
$L\times L$ lattice, we traverse the $4(L-1)$ boundary sites and remove each spin with probability 
$p=1/3$ or couple a new spin to it with the same $p$, doing nothing with probability $1-2p$. 
B) With spins randomly occupying the sites of an open $L\times L$ lattice with probability $p$, 
we identify the largest cluster. C) Starting from a single occupied site on an infinite 
lattice, we add neighbors to it with probability $p$, filling the neighbors of added spins with this same 
probability, until a cluster of a desired size $N$ has formed. We here only consider the case $p=0.7$ and $0.6$ 
for $B$ and $C$ clusters, respectively. Since we want to isolate the effects of edges, in all cases we also fill any 
internal vacancies, so that all sites within a single boundary are magnetic. The $C$ clusters  are close to the 
percolation point ($p\approx 0.59$) and are therefore very (unrealistically) irregular in shape, thus serving 
as an extreme case. 

In all cases, we here initially consider only clusters with the same number of spins on both sublattices. 
This is done in order to avoid a trivial enhancement of the susceptibility originating just from the fact
that clusters with sublattice imbalance have ground states with non-zero total spin. On the other hand, 
in real systems, sublattice imbalanced clusters will of course be present, and internal 
vacancies or defects can be expected as well. Our calculations here are intended to better isolate the 
effects solely due to the edges. We will also investigate the effects of the fluctuation of the
total ground-state spin in the latter part of this paper.

\begin{figure}
\includegraphics[width=6.5cm, clip]{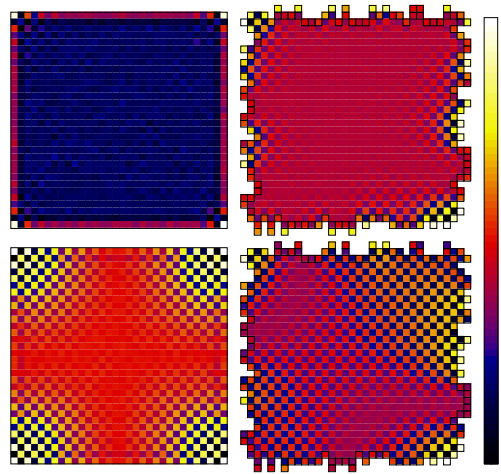}
\vskip-1mm
\caption{(Color online) Local magnetic response for two systems; an intact $32\times32$  
lattice (left) and a rough-edge $A$ cluster (right). The upper and lower graphs are for 
$\beta=2$ and $\beta=10$, respectively. The intensity color-codings correspond to 
lowest-highest response according to: $0.056 \leq \chi_l \leq 0.186$ (left, upper), 
$-0.19 \leq \chi_l \leq 0.47$ (right, upper), $-0.087 \leq \chi_l \leq 0.177$ (left, lower),
$-1.8 \leq \chi_l \leq 2.9$ (right, lower).}
\label{clusters}
\vskip-2mm
\end{figure}

Fig.~\ref{clusters} shows the local susceptibility landscape of an intact open lattice as well 
as a rough-edge cluster of type $A$. For the smooth-edge cluster, the corners have the largest response. 
Because of this, in the presence of a weak external field the neighbors of the corners would experience an effectively
negative field, and, thus, the susceptibility of these spins is negative at low temperatures. This 
kind of staggered susceptibility pattern continues away from the corners on a length-scale which 
should be related to the exponentially divergent correlation length \cite{chn}. A similar effect 
around an isolated impurity has been studied in great detail previously \cite{anfuso}. Note that
the low-$T$ suppression of the edge susceptibility found in \cite{hoglund2} refers to the uniform component
(average over spins at a fixed distance from the edge), whereas the staggered response is enhanced
at the edge.

Our main interest here is in the response at and close to a rough edge, where strong staggered susceptibility patterns 
can appear around some spins for the same reasons as discussed above. Because of interference among several
randomly located high-response centers, these effects are amplified in some regions and damped in others,  as 
seen in the type $A$ cluster in Fig.~\ref{clusters}, leading to a complex behavior with much larger 
fluctuations in local susceptibilities than in the intact open system.

To characterize the susceptibility distribution as a function of the cluster size, we define the average radius of
a cluster $c$ with respect to the site closest to the ``gravitational center'' $\mathbf r_0^c$, and further average 
this radius over different cluster realizations;  
\begin{equation}
\langle R \rangle =\sqrt{\frac{1}{n_c}\sum_c\frac{1}{N_e^c} \sum_{e}({\mathbf r_{e}^c}-\mathbf r_0^c)^2}.
\end{equation}
Here $\mathbf r_e^c$ is the position of an edge spin and $N_e^c$ is the number of edge spins in cluster $c$. 
 
\begin{figure}
\includegraphics[width=7.8cm, clip]{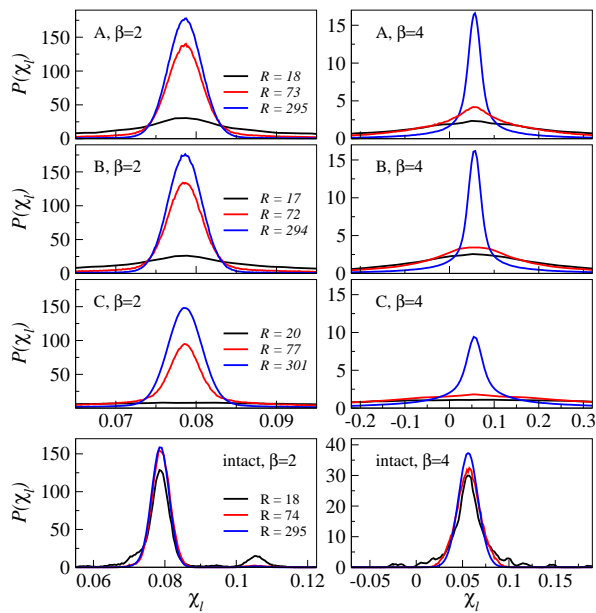}
\vskip-2mm
\caption{(Color online) Local susceptibility distribution for clusters of types $A$, $B$, $C$, as well as intact open lattices, 
all at two different inverse temperatures; $\beta=2$ and $\beta=4$. Note that different scales are used for the intact 
lattices (of sizes $L=32$, $128$, $512$ corresponding to the average radii $R$ indicated). The results for A, B, and C
clusters are averages over 200-1000 different random realizations.}
\label{chihist}
\vskip-2mm
\end{figure}

Fig.~\ref{chihist} shows histograms of the local susceptibility (corresponding to the NMR profile) at inverse temperatures 
$\beta=2$ and $4$. Note that, in the context of cuprates, these temperatures are rather high, with $\beta=4$ corresponding 
approximately to room temperature (close to $T_{\rm N}$ for La$_2$CuO$_4$). NMR experiments have, however, in the past
been carried out even up to $T=900$ K \cite{imai}, and the behavior there is still in good agreement with the 
2D Heisenberg model \cite{sandviknmr}. For all clusters, the shift distribution
is much wider at $\beta=4$ than at $\beta=2$. At the higher temperature ($\beta=2$), the bulk susceptibility rapidly starts 
to dominate as the cluster size is increased. At $\beta=4$ there are, however, still very significant broadening 
effects for clusters as large as $R\approx 300$. Experimentally, for fine-powdered cuprate samples, one would expect clusters roughly of 
this size or somewhat larger \cite{takigawacomm}. For small intact $L\times L$ clusters at $\beta=2$, the edges give rise to a separate
large peak at $\chi_l \approx 0.1$. There is also a much smaller peak due to the corners at higher $\chi_l$ (not seen in the figure). 
At $\beta=4$ several smaller peaks originate from frames of spins at and close to the edge  \cite{hoglund2}. These effects 
persist for larger clusters but become difficult to discern because of their low relative weight. Overall, the random edges lead to 
much broader peaks.

To quantify the width of the susceptibility distribution, we calculate the standard 
deviation with respect to the average local susceptibility over all clusters,
\begin{equation}
\sigma_{\chi} = \sqrt{\frac{1}{\sum_{c} N_c }\sum_{c}\sum_{r=1}^{N_c} (\chi_l^c (r)-\langle \chi_l \rangle)^2} ,
\end{equation}
where $c$ labels the individual clusters, $N_c$ is the number of occupied sites of the clusters, and $r$ is a site label 
for these magnetic sites; thus, $\chi_l^c (r)$ is the local magnetic susceptibility of site $r$ on cluster $c$. Its average over all 
spins of all clusters is denoted $\langle \chi_l \rangle$. 

\begin{figure}
\includegraphics[width=5.9cm, clip]{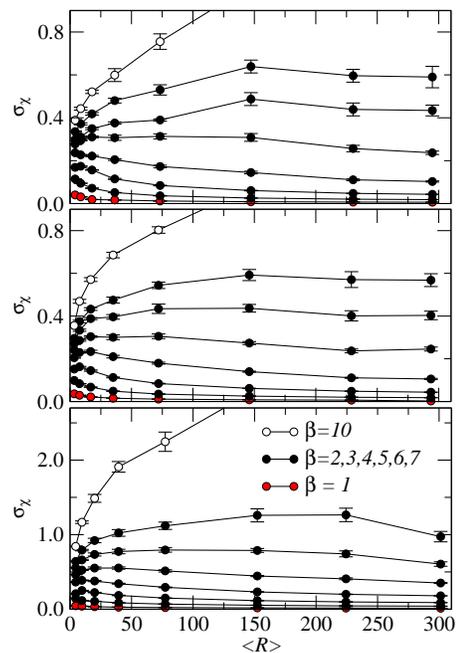}
\vskip-2mm
\caption{(Color online) Standard deviation of the local susceptibility distribution for the three types of clusters
as functions of the effective radius $\langle R \rangle$ at several inverse temperatures.}
\label{chisigma}
\vskip-2mm
\end{figure}

Fig. \ref{chisigma} shows results as a function of the average cluster size $\langle R\rangle$ for inverse temperatures in the range $\beta 
\in [1,10]$. In all cases, the line width first grows with the cluster size and then decreases. For any finite $\beta$, as $\langle R\rangle \to \infty$
we must have $\sigma_{\chi} \to 0$, as the edge-to-bulk ratio vanishes for an infinite cluster and the edge effects can only extend a finite 
distance away from the edge at any $T>0$. The initial increases in $\sigma_x$ reflect the tails of the distributions, which only 
develop fully for large clusters. The tails are very significant at low temperatures, i.e., the rough edges influence the response far inside 
the bulk of the clusters (reflecting the exponentially divergent correlation length for $T\to 0$ \cite{chn}). At the lowest temperature ($\beta=10$) 
the line widths still increase for the largest clusters we have studied. Comparing the line-widths of the three types of clusters, we see that 
type $C$ always produces the broadest lines (for given temperature and cluster radius). These cluster are also the ones that are the geometrically 
most rough ones, being constructed close to the geometrical percolation point. Real cuprate clusters are likely not as rough. Note, however,
that even clusters of type $A$, for which the geometrical edge disorder is very shallow, lead to distributions only a factor of $\approx 2$ narrower.

For all the disordered clusters with $\langle R \rangle \approx 300$, the line widths exceed the average Knight shift for $\beta = 3\sim 5$; 
relevant for cuprates in their paramagnetic state. At $\beta=4$, $A$ clusters with $\langle R \rangle=295$ give 
$\sigma_{\chi}=0.103$, while the average shift is $\langle \chi_l \rangle = 0.0576$, type $B$ clusters with  $\langle R \rangle=294$ give 
$\sigma_{\chi}=0.105$ and $\langle \chi_l \rangle =0.0571$, and type $C$ clusters with $\langle R \rangle =301$ have $\sigma_{\chi}=0.179$, 
$\langle \chi_l \rangle =0.0622$. Line broadening at this level should be clearly visible experimentally (although the long tails of the 
distributions may be partially drowned by experimental noise). 

\begin{figure}
\includegraphics[width=6.2cm,clip]{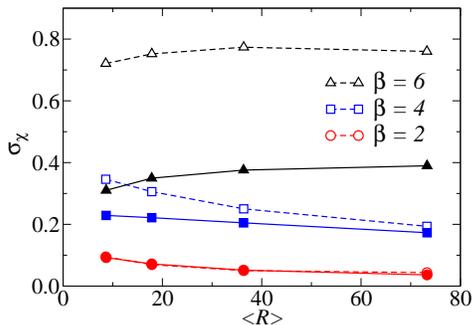}
\vskip-2mm
\caption{(Color online) Size dependence of the line width $\sigma_{\chi}$ for type $A$ clusters constructed under the condition of 
equal numbers of spins on the two sublattices (filled symbols), and without this restriction (open symbols).}
\vskip-2mm
\label{chinoneq}
\end{figure}

We now discuss the case of averages over clusters without the restriction of equal numbers of spins on both sublattices. 
Fig.~\ref{chinoneq} shows the line widths for the two ensembles of type $A$ clusters. At high temperatures, the line widths are almost equal
(indistinguishable in the figure at $\beta=2$) widths. At lower temperatures, the unrestricted ensemble gives significantly broader lines, 
however (about a factor of two  at $\beta=6$). For very large clusters we expect no differences, because the relative fluctuation of the 
difference in sublattice occupation is $\sim 1/\sqrt{N_c}$. The convergence of the two ensembles with increasing $\langle R\rangle$ is seen clearly 
for $\beta=4$. Based on these results, we expect the contributions from the sublattice imbalance to be small relative to the dominant 
boundary-roughness effects in the size and temperature regimes most relevant to cuprates.

In summary, we have shown that significant NMR line-broadening should be expected due to rough edges of nano-scale antiferromagnetic clusters. 
In the temperature range of relevance, e.g., for the cuprates in their paramagnetic state, the line widths produced by all the 
cluster types studied are larger than the average Knight shift, even for clusters several hundred lattice constants wide. 
The effects should therefore be experimentally observable in fine-powdered samples. 

In light of these results, it may seem surprising that no anomalous broadening has been noted in experiments so far (perhaps
suggesting that the edges actually are rather smooth). It should therefore be interesting to systematically search for edge effects in NMR and
other experiments. In order to make the NMR broadening more prominent and investigate the size dependence, it would then be desirable to prepare 
powder samples with very small clusters, down to $\approx 100$ lattice spacings across. Current techniques (which are normally not intended to produce 
extremely small clusters) likely give clusters roughly an order of magnitude larger \cite{takigawacomm}, but other methods could perhaps reach 
smaller sizes. Systematical studies of antiferromagnetic correlations and response functions in nano-scale systems may give further insights into 
the microscopic interactions and collective quantum phenomena in strongly-correlated electron systems. 

We would like to thank Masashi Takigawa for very useful discussions and suggestions. The work of WG was supported by the NSFC under Grant No.~10675021, 
and by the High Performance Scientific Computing Center of the Beijing Normal University. WG also acknowledges hospitality and financial support 
from the Condensed Matter Theory Visitors Program at Boston University. The work of AWS was supported by the NSF under Grant No.~DMR-0803510. 

\null

\end{document}